\documentclass[aps,pra,groupedaddress,floatfix,showpacs,twocolumn]{revtex4-1}
\usepackage{amsmath}
\DeclareMathOperator{\tr}{tr}
\DeclareMathOperator{\imag}{Im}
\usepackage{graphicx}
\usepackage{bm}
\usepackage{epstopdf}
\usepackage[caption=false]{subfig}
\usepackage{microtype}
\usepackage{placeins}
\usepackage{braket}
\usepackage{xcolor}

\begin{document}

\title{Probabilistic Hysteresis from a Quantum Phase Space Perspective}

\author{Ralf~B\"urkle$^{1}$}
\email{rbuerkle@rhrk.uni-kl.de}
\author{James R.~Anglin$^{1}$}

\affiliation{$^{1}$\mbox{State Research Center OPTIMAS and Fachbereich Physik,} \mbox{Technische Universit\"at Kaiserslautern,} \mbox{D-67663 Kaiserslautern, Germany}}

\date{\today}

\begin{abstract}
\emph{Probabilistic hysteresis} is a manifestation of irreversibility in a small, dissipationless classical system [Sci. Rep. 9, 14169]: after a slow cyclic sweep of a control parameter, the probability that an initial microcanonical ensemble returns to the neighborhood of its initial energy is significantly below one. A similar phenomenon has recently been confirmed in a corresponding quantum system, when the particle number $N$ is not too small. Quantum-classical correspondence has been found to be non-trivial in this case, however; the rate at which the control parameter changes must not be extremely slow and the initial distribution of energies must not be too narrow. In this paper we directly compare the quantum and classical forms of probabilistic hysteresis by making use of the Husimi quantum phase space formalism. In particular we demonstrate that the classical ergodization mechanism, which is a key ingredient in classical probabilistic hysteresis, can lead to a breakdown of quantum-classical correspondence rather than to quantum ergodization. Such quantum failure of ergodization leads to strong quantum effects on the long-term evolution even when the quantum corrections in the equations of motion, which are proportional to  $1/N$, would naively seem to be small. We also show, however, that quantum ergodization can be restored by averaging over energies, so that for not-too-narrow initial energy width and not-too-slow parameter change the classical results are recovered after all at large $N$. Finally we show that the formal incommutability of the classical and adiabatic limits in our system, which is responsible for the breakdown of quantum-classical correspondence in the quasi-static limit, is due to macroscopic quantum tunneling through a large energetic barrier. This explains the extremely slow sweep rates needed to reach the quantum adiabatic limit that were reported in our previous work. The formal incommutability therefore has no consequences for any realistically slow sweeps unless $N$ is quite small ($N \lesssim 20$). 

\end{abstract}

\maketitle
  
\section{Introduction}
\subsection{The microscopic onset of irreversibility}
Recently the phenomenon of \emph{probabilistic hysteresis} has been introduced \cite{dimer, trimer, quantum_dimer_I}: A slow, cyclic sweep of an external parameter can lead to a state that is very different from the initial state, even though the external parameter is tuned back to its initial value. This phenomenon can be interpreted as the microscopic analogue of macroscopic irreversible processes as e.g. cooking an egg and then cooling it down again to the initial temperature. As everyday experience tells us, the initial raw egg will not be recovered. From the adiabatic theorem, however, one would rather expect the system to follow a stationary state as long as the parameter sweep is sufficiently slow, so that it would return to its initial state again when the cycle of the sweep is complete. While one might not expect the adiabatic theorem to apply to macroscopic systems, which typically have many low-frequency degrees of freedom, it should be possible to attain the adiabatic limit in sufficiently small systems. The possibility of this egg-cooking kind of irreversibility even in a dissipationless microscopic system is therefore surprising.

In \cite{dimer, trimer} we have shown how this irreversibility can occur due to the crossing of a separatrix in phase space, where adiabaticity breaks down even for arbitrarily slow change of the control parameter. This is the case because the orbital period diverges at the separatrix and so the criterion for the validity of the adiabatic theorem can never be met, no matter how slow the variation of the external parameter might be. In the integrable system of \cite{dimer} this results in a finite probability to either return to the initial state or to a state with much higher energy (thus \emph{probabilistic} hysteresis); in the chaotic system of \cite{trimer} the return probability is typically very close to zero. 

\subsection{Quantum-classical correspondence}
In our previous work this phenomenon was identified in two specific models for trapped ultracold atom systems, namely the Bose-Hubbard dimer \cite{dimer} and trimer \cite{trimer}, in a semiclassical mean-field approximation. Although in classical Hamiltonian evolution adiabaticity can fail even in the quasi-static limit of infinitely slow sweep rate, this quasi-static limit \emph{must} be adiabatic under fully quantum-mechanical evolution, because quantum energy level splittings always remain non-zero in this system (i.e. there can never be any exact degeneracies). This means that the classical and adiabatic limits do not commute, as has been noted in the literature \cite{Wu,Berry}. In the true quantum adiabatic limit, therefore, the sweep process is necessarily fully reversible and hysteresis is absent. 

How slow does the sweep have to be, though, to actually reach this true quantum adiabatic limit? It turns out that even for quite small total particle numbers $N=\mathcal O(10)$, and for any remotely realistic values of the other system parameters, the sweep time has to be quite unrealistically slow: anywhere from several years up to many times the age of the universe \cite{quantum_dimer_I}. In \cite{quantum_dimer_I} we used Landau-Zener theory to describe the cyclic sweep process in the quantum version of the Bose-Hubbard dimer system that we had previously studied in \cite{dimer}, and compared a wide range of different sweep rates. We found that for not too small total particle numbers and a broad range of sweep rates the basic classical picture of two qualitatively different final states being reached probabilistically is recovered quantum mechanically. The quantum probability to recover the initial state (the \emph{return probability}), however, was found to depend sensitively on the sweep rate; it oscillates around the constant classical quasi-static value, with finite frequency and significant amplitude, even for very large particle numbers. We confirmed numerically that this non-classical oscillation of the return probability with sweep rate disappears only if, in addition to having large $N$, the initial quantum state is not a single energy eigenstate but a mixed state with a sufficient energy width.

\subsection{A quantum phase space picture}
These results may have shed some light on the role of quantum mechanics in the microscopic onset of irreversibility, but they have not clarified that role as well as one might wish, because quantum and classical probabilistic hysteresis have been described in such different terms. In the classical system the process is quite clear in phase space \cite{dimer}; it combines incompressible phase-space flow under Liouville's theorem, topological change of energy surfaces as they merge and separate, and effective ergodization through very fine swirling of initially coarse distributions. The return probability in the slow-sweep limit could even be computed analytically by applying Kruskal's theorem \cite{dimer}. The quantum phenomenon was in contrast described in terms of a sequence of Landau-Zener transitions between adiabatic quantum many-body eigenstates. The clear classical explanation for probabilistic hysteresis was hard to discern in this sequence, and the recovery of the classical return probability through a combination of many Landau-Zener transitions seemed to be a sheer numerical conspiracy. 

To gain more insight into quantum irreversibility we therefore turn in this paper to a phase space representation of quantum dynamics, since the classical phase space picture is clear and quantum phase space methods have often proven to be very useful in understanding quantum-classical correspondence \cite{Polkovnikov_TWA, Witthaut2, Weinbub, Takahashi, Mahmud, Torres-Vega}. In particular we will use the Husimi quasi-probabiltiy function as a representation for the quantum states in phase space and compare its full quantum evolution to the semiclassical Truncated Husimi approximation \cite{Vermersch, Witthaut2, Trimborn}. While this quantum phase space description brings us closer to an analytical understanding of the purely numerical results of \cite{quantum_dimer_I} for the return probability, it also shows why quantum-classical correspondence can still break down for large total particle numbers, resulting in strong quantum effects. This is especially surprising because a naive argument suggests that the quantum correction term scales like $1/N$ (see Sec.~\ref{Husimi}). Furthermore the Husimi description allows us to distinguish the two qualitatively different effects of quantum noise and quantum interference. Finally, the Husimi phase space description will provide an intuitive understanding of why the true quantum adiabatic limit is so extremely hard to reach.

\subsection{The quantum Bose-Hubbard dimer}
The two-site Bose-Hubbard ``dimer'' system has been realized experimentally in ultracold atom systems \cite{Oberthaler1, Oberthaler2}; well before this achievement its theoretical study had already been extensive. Several previous works \cite{Wu2, Zobay, Liu, Wu4, Jona-Lasinio,Yang} have even specifically addressed slow parameter sweeps in this model. As discussed in more detail in \cite{quantum_dimer_I}, however, these earlier papers have used some typically quantum terminology (such as ``tunneling probability'') but have actually been restricted to the classical, mean-field version of the problem, and so do not really bear on our current topic.

In contrast to the two-state nonlinear Schr\"odinger evolution of mean-field theory, the full $N$-particle quantum many-body system of the Bose-Hubbard dimer concerns an $(N+1)$-component wave function evolving under a \emph{linear} Schr\"odinger equation. This problem has been studied for a single non-cyclic sweep in \cite{Korsch, Trimborn3, Wu, Chen}; it has been shown that the many-body Landau-Zener probability for a diabatic transition between the quantum levels goes to zero in the adiabatic limit of infinitely slow sweep rate, in accordance with the quantum adiabatic theorem. While this means that the mean-field and adiabatic limits do not commute, as mentioned above, it has also been demonstrated \cite{Korsch,Trimborn3,Wu} that for a fixed slow but finite sweep rate the ``Landau-Zener probability'' (i.e. the ratio of $\braket{n_2}$ to $\braket{n_1}$)  approaches the mean-field value quite rapidly with increasing $N$, with good quantum-classical correspondence already for $N=\mathcal O (10)$. 

As we have numerically demonstrated in \cite{quantum_dimer_I}, however, quantum-classical correspondence is more subtle than this for the phenomenon of probabilistic hysteresis, for two main reasons. Firstly, probabilistic hysteresis can occur for a finite range of initial states, not only the initial ground state, and the simple correspondence of the ground state turns out to be a special case. Secondly the scenario of probabilistic hysteresis includes a second, backward sweep, which begins from the excited state that was created by the forward sweep even when the initial state was the ground state. All cases of probabilistic hysteresis therefore turn out to involve significant quantum interference effects which are not captured by the semiclassical approximation and which persist even in the limit of very large $N$.

\subsection{Structure of the paper}
The rest of the paper is organized as follows: In Sec.~II we present the Hamiltonian and the sweep protocol that we will study and we also briefly review the semiclassical description of the sweep process. We will show how hysteresis and the finite return probability in the classical adiabatic limit can be understood by considerations in phase space. In Sec.~III we first introduce appropriate coherent states and the Husimi function. We then show how the semiclassical evolution in phase space and the evolution of the quantum Husimi function are related. We demonstrate that the classical ergodization mechanism fails to produce quantum ergodization, but instead induces the breakdown of quantum-classical correspondence, and is thus responsible for the strong quantum effects that occur, for single initial energy eigenstates, even at large total particle numbers. We also show how ergodization in the quantum system can be restored by a different mechanism so that for a finite initial energy width the semiclassical results are recovered after all. After a brief discussion of the entropy generated in the classical and quantum sweep process we show in Sec.~IV that the quantum adiabatic limit, in which the return probability is always one, can be understood as macroscopic quantum tunneling of a large number of atoms through the separatrix energy barrier and is therefore exponentially slow. We then proceed to Sec.~V where we summarize our main results. 

\section{Setup and semiclassical description}
\subsection{Setup}
Our system is the two-mode Bose-Hubbard system with attractive interaction $U<0$ and tunneling rate $\Omega$. The two modes have a time-dependent energy detuning $\Delta(t)$, which will be our control parameter. The system Hamiltonian therefore reads
\begin{equation}
\hat H = -\frac{\Omega}{2}(\hat{a}^{\dagger}_{1}\hat{a}_{2}+\hat{a}^{\dagger}_{2}\hat{a}_{1})+\frac{U}{2}(\hat{n}_{1}^{2}+\hat{n}_{2}^{2})+\frac{\Delta(t)}{2}(\hat{n}_{1}-\hat{n}_{2}), \label{eq:H}
\end{equation}
where the bosonic operators $\hat a_{1,2}^{\dagger}$ ($\hat a_{1,2}$) create (destroy) a boson in the respective mode 1 or 2 and the number operators $\hat n_{1,2}=\hat a_{1,2}^{\dagger} \hat a_{1,2}$ are defined as usual. In this paper we choose units such that $\hbar=1$ and measure $\Delta$, $U$, energy and time in units defined by $\Omega$. The total particle number operator $\hat N=\hat n_1+\hat n_2$ commutes with the Hamiltonian, so that the total particle number given by its eigenvalue $N$ is conserved.

Our protocol consists of slowly ($T\gg \Omega^{-1}$) sweeping the energy detuning $\Delta(t)$ from a negative value $\Delta_I$ at the initial time $t=-T$ to the larger value $\Delta_0$ at $t=0$ (forward sweep) and then back again to $\Delta_I$ at the final time $t=+T$ (backward sweep):
\begin{equation}
\Delta(t)= \Delta_{I}\frac{|t|}{T}+\Delta_{0}\left(1-\frac{|t|}{T}\right), \qquad \Delta_0>\Delta_I.
\label{eq:sweep}
\end{equation} 
We will study the evolution of a quantum state during this cyclic sweep; as initial states we will choose either a low-lying instantaneous energy eigenstate of the Hamiltonian with fixed $\Delta=\Delta_I$, or else a narrow microcanonical ensemble of such eigenstates. We then ask the question: With what probability is the initial state recovered at the final time, after the slow forward-and-back cycle of $\Delta$?

\subsection{Semiclassical picture}
\label{semiclassical}
The semiclassical description \cite{dimer} of the quantum sweep process is obtained by evolving an ensemble of initial phase space points that represent the initial quantum state under the mean-field equations of motion (Truncated Wigner or Truncated Husimi approximation). These equations of motion can be derived from the mean-field Hamiltonian
\begin{equation}
H=-\Omega \sqrt{p_0^2-p^2} \cos(q)+U\left(p_0^2+p^2\right)+\Delta(t) p,
\label{eq:H_cl}
\end{equation}
where $(q,p)$ are canonical coordinates representing the relative phase and particle imbalance, respectively \cite{dimer}. As always in these types of mean-field systems the system behavior can be characterized by the single parameter $u=UN/\Omega$ since Hamiltonians for different total particle numbers but same $u$ can be mapped onto each other by trivial rescaling.

For the following discussion let us assume as our initial state a microcanonical ensemble (i.e. a complete thin shell of fixed energy) at the initial detuning $\Delta_I$. Our reasoning may then be applied to arbitrary phase space distributions where the probability depends only on energy, by viewing them as consisting of a large number of narrow microcanonical ensembles. We sample this initial phase space density with a finite number of points and evolve each point under the mean-field equations of motion derived from Eq.~(\ref{eq:H_cl}). Since our sweep is slow compared to $\Omega^{-1}$ the classical adiabatic theorem can be applied unless the orbital period deviates significantly from $\Omega^{-1}$. This is not the case in the subcritical case $|u|<1$ (which means $u>-1$ for our attractive negative $U$) and so the action of each trajectory is an adiabatic invariant \cite{Goldstein}. This means that the orbits deform and their energies change during the forward sweep, but in a way that keeps their enclosed phase space area constant. During the backward sweep the same deformation happens in reverse and the initial and final ensemble coincide; consequently the return probability is one.

\begin{figure*}
\centering
\subfloat[Initial state]{\includegraphics[width=.32\textwidth]{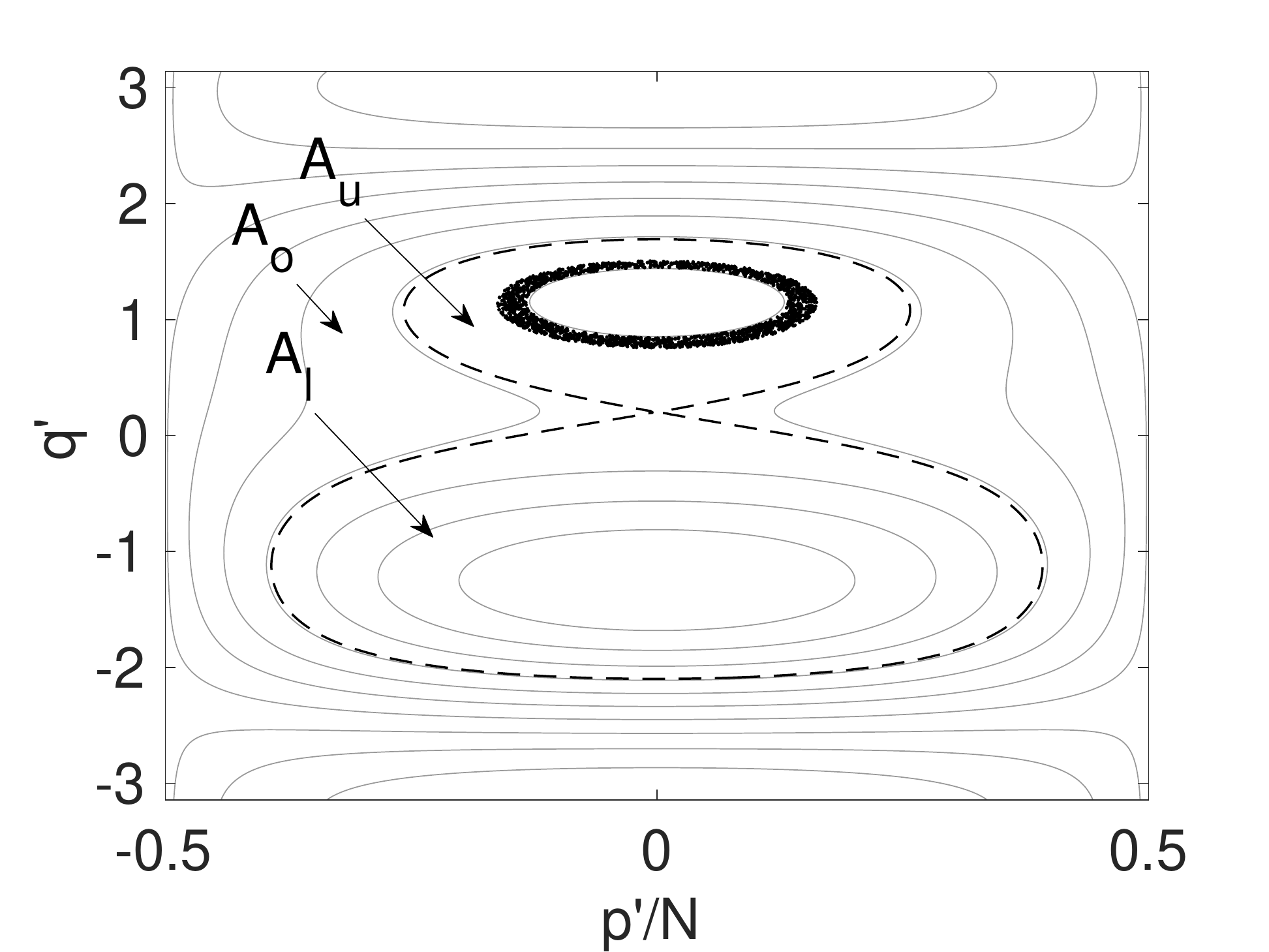}}
\subfloat[End of forward sweep]{\includegraphics[width=.32\textwidth]{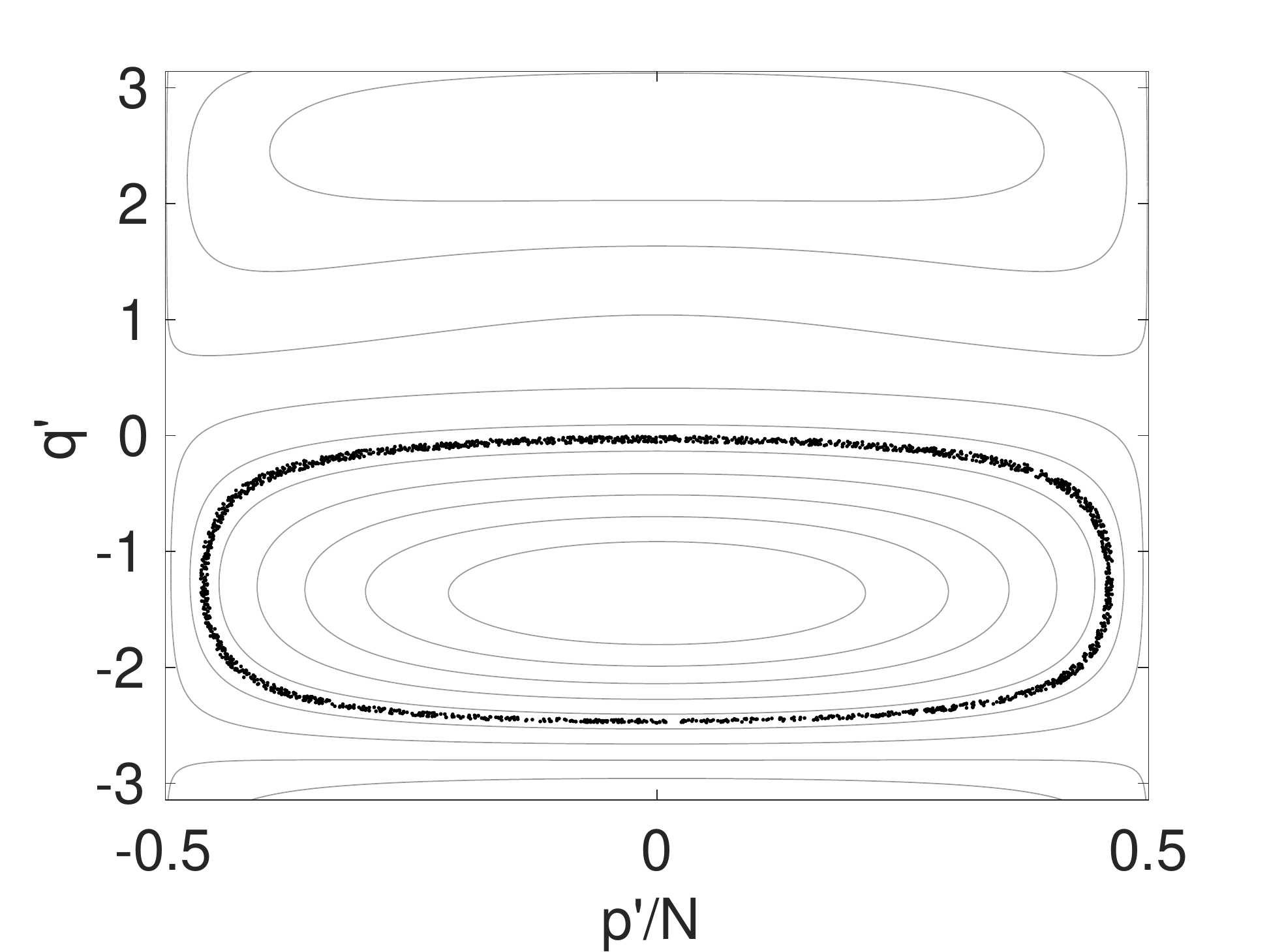}}
\subfloat[Final state]{\includegraphics[width=.32\textwidth]{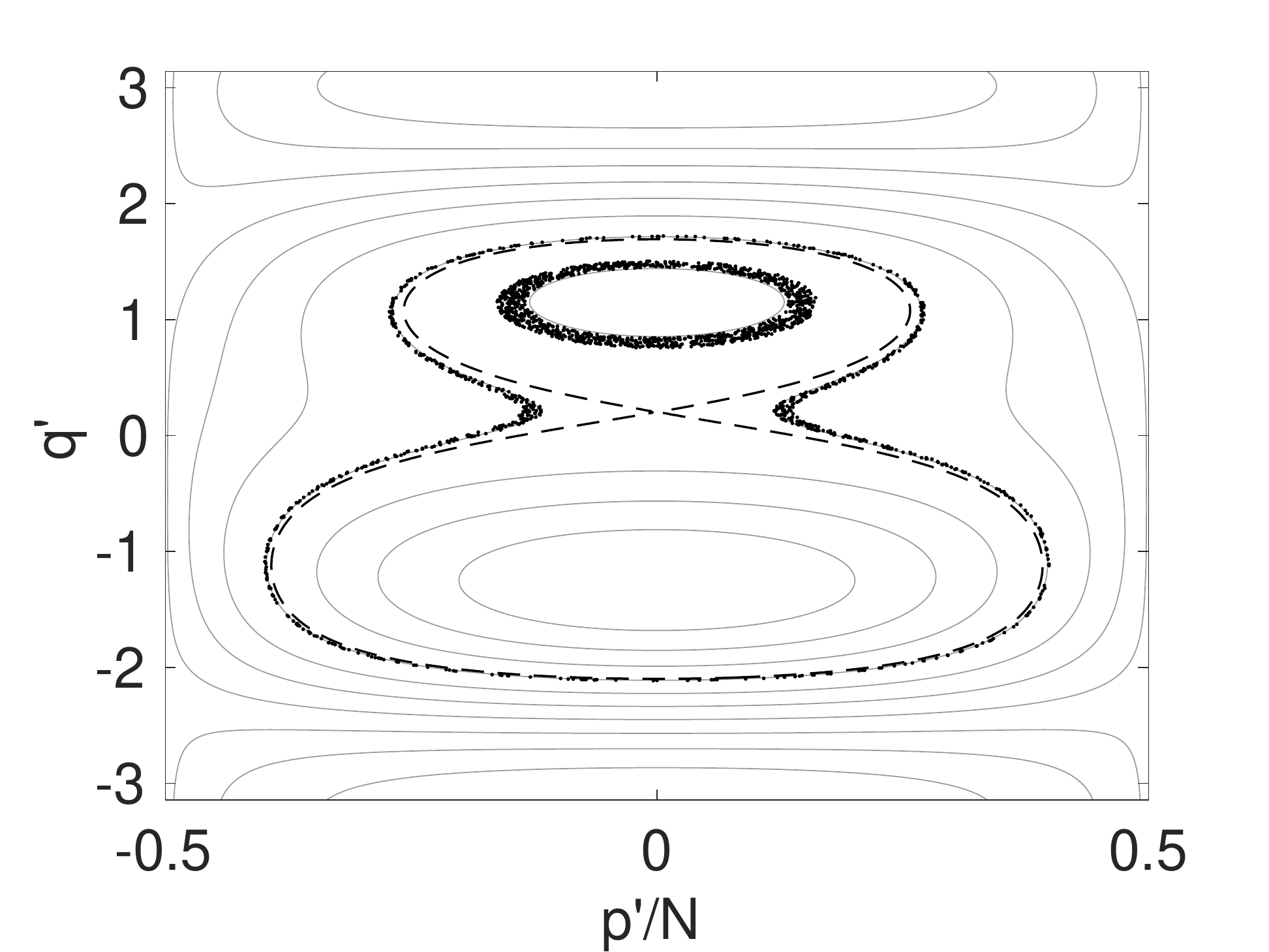}}
\caption{(Reproduced from Fig.~1 of Ref.~\cite{quantum_dimer_I}.) Evolution of a classical ensemble consisting of 2000 points (black dots) in phase space. The gray lines show adiabatic energy contours; the dashed black line is the separatrix that divides phase space into the regions $A_u$, $A_l$ and $A_o$. Because adiabaticity breaks down when the separatrix is crossed between (a) and (b) and between (b) and (c), only a finite fraction of the ensemble returns to the initial energy shell, so that probabilistic hysteresis occurs. For a clearer graphical presentation we have chosen the canonical coordinates $q'=\arctan\left(p/\left(\sqrt{p_0^2-p^2}\cos(q)\right)\right)$, $p'=-\sqrt{p_0^2-p^2}\sin(q)$ here. }
\label{fig:classical_phase_space}
\end{figure*}

In the supercritical case $|u|>1$ ($u<-1$), on the other hand, there is an unstable fixed point in a certain $\Delta$ range. The energy contour running through this unstable fixed point, the \emph{separatrix}, divides the phase space into three mutually exclusive regions that we label $A_u$, $A_l$ and $A_o$ (see Fig.~\ref{fig:classical_phase_space}). Here and in the rest of the paper we consider $u=-3$ but other supercritical values give similar results.
When the separatrix first forms the entire ensemble resides within the upper lobe of the separatrix $A_u$---see Fig.~\ref{fig:classical_phase_space}(a). However, this separatrix lobe shrinks during the forward sweep so that at some point it meets the ensemble. As the separatrix approaches the ensemble, the orbital period of the ensemble members grows and finally diverges, since the orbital period associated with the separatrix itself is infinite. This means that the adiabatic theorem no longer holds at the separatrix, no matter how slow the parameter sweep may be. As a result the trajectories change their enclosed actions by leaving the upper separatrix lobe and moving into the only growing phase space region, namely the lower separatrix lobe. The continued growth of this lower lobe then means that the separatrix expands away from the ensemble, and so adiabaticity is restored to the ensemble again for the rest of the forward sweep (Fig.~\ref{fig:classical_phase_space}(b)). 

During the backward sweep, then, adiabaticity continues to hold until the now-shrinking lower lobe of the separatrix $A_l$ hits the ensemble, at the same value of the detuning at which the separatrix crossing occurred in the forward sweep. Adiabaticity breaks down again, just as during the forward sweep, but now the upper lobe $A_u$ and the outside region $A_o$ are both growing, so that the members of the ensemble can go into either of these phase space regions. Kruskal's theorem (see \cite{LC} and references therein), which is derived from Liouville's theorem, gives the proportion of the parts of the ensemble going to the upper lobe $A_u$ and the outside region $A_o$. Since adiabaticity holds once again after the separatrix has been crossed for the second time, the part of the ensemble that went to the upper separatrix lobe ends up in the same energy shell in which it started initially. The rest of the ensemble, however, ends up with a much higher energy than it had initially, so that the final state consists of two well-separated sub-ensembles (Fig.~\ref{fig:classical_phase_space}(c)). We then define the return probability $P_{\mathrm{ret}}$ as the fraction of ensemble members that returned to the initial energy shell.

It is important to note that in the semiclassical system a well-defined quasi-static limit exists, even though adiabaticity is necessarily broken at some point, in the sense that the return probability settles quickly to the value predicted by Kruskal's theorem once the sweep rate falls below a certain finite range. Further reducing the sweep rate does not alter the return probability further. 

Depending on the system parameters and on the energy of the initial ensemble, the return probability of a finite-width ensemble can range between almost zero and almost one. The essentially-zero return probability that is familiar from macroscopic systems for all initial conditions is not present in our simple integrable system; it can, however, be realized in a similar trimer system that allows chaotic dynamics \cite{trimer}. The general relationships between quantum chaos, quantum ergodicity, and irreversibility \cite{Zhang2,Peres} clearly require much further study; as a first step toward understanding probabilistic hysteresis as a particularly simple form of microscopic irreversibility, here we will consider only the quantum version of the integrable dimer system, and leave the quantum version of the non-integrable trimer for future work.

\section{Quantum Phase Space Picture}
\subsection{Husimi function}
\label{Husimi}
The classical phenomenon of probabilistic hysteresis concerns ensemble evolution in phase space, rather than the motion of individual phase space points. The classical evolution of the phase space density $\rho(q,p)$ is given by the Liouville equation, which for the Hamiltonian (\ref{eq:H_cl}) reads
\begin{equation}
\begin{split}
\dot \rho&=\frac{\partial \rho}{\partial p} \frac{\partial H}{\partial q}-\frac{\partial \rho}{\partial q} \frac{\partial H}{\partial p}\\
&=\Omega \sqrt{p_0^2-p^2} \sin(q) \frac{\partial \rho}{\partial p}\\
&-\left(2Up+\Delta+\frac{\Omega p}{\sqrt{p_0^2-p^2}}\cos(q)\right) \frac{\partial \rho}{\partial q}.
\end{split}
\label{eq:Liouville}
\end{equation}
In order to compare directly with this classical evolution, therefore, we must also formulate the quantum evolution in phase space, using a quantum quasi-probability function. In particular we use the Husimi function $Q$ \cite{Husimi}, which is defined as the probability to find the quantum system in a coherent state $\ket \Gamma$
\begin{equation}
Q(\Gamma,t)=|\braket{\Gamma|\psi(t)}|^2=\braket{\Gamma|\hat \rho(t)|\Gamma}.
\label{eq:Husimi_def}
\end{equation} 
The definition of the Husimi function is motivated by the fact that coherent states are the most localized quantum states in phase space, in the sense that they minimize the uncertainty product of the phase space variables, so that they are the closest quantum analogues of classical phase space points. Despite this intuitive meaning of the Husimi function, one cannot simply interpret $Q$ as a probability distribution in phase space, because it does not necessarily give the correct marginal distributions if one of the phase space variables is integrated out. The Husimi function does provide a complete description of the quantum state, however, in the sense that all information about the quantum state can in principle be extracted from $Q$.

Using the von Neumann equation for the evolution of the density operator $\hat \rho$, the time evolution of the Husimi function is given by
\begin{equation}
\begin{split}
\dot Q&=\braket{\Gamma|\dot {\hat \rho}|\Gamma}=\tr \left(\dot{\hat \rho} \ket \Gamma \bra \Gamma\right)\\
&=\tr\left(i \hat \rho \hat H \ket \Gamma \bra \Gamma-i\ket \Gamma \bra \Gamma \hat H \hat \rho \right).
\label{eq:Qdot}
\end{split}
\end{equation}
Because the symmetry group of the Bose-Hubbard dimer is $SU(2)$ and it is therefore equivalent to a spin system with $s=p_0=N/2$ \cite{Schwinger, Sakurai}, the appropriate generalized coherent states are the so-called SU(2) coherent states \cite{Arecchi, Gilmore, Narducci, Narducci2, Trimborn, Witthaut2}
\begin{equation}
\begin{split}
\ket{\Gamma}&=\ket{\theta,\phi}\\&
=\sum_{n=0}^N \sqrt{{N}\choose {n}} \left(\cos \frac{\theta}{2}\right)^{n} \left(\sin \frac{\theta}{2} e^{i\phi}\right)^{N-n} \ket{n,N-n},
\end{split}
\end{equation}
where $(\theta,\phi)$ are the angles in a spherical coordinate system, namely the Bloch sphere. The classical canonical coordinates $(q,p)$, which have already been used in Sec.~II, are given by $q=\phi$ and $p=N/2 \cos(\theta)$; they map the Bloch sphere onto a flat phase space.

Defining the angular momentum operators
\begin{equation}
\begin{split}
\hat L_x&=\frac 12 \left(\hat a_1^{\dagger} \hat a_2+ \hat a_2^{\dagger} \hat a_1\right),\\
\hat L_y&=\frac i2 \left(\hat a_2^{\dagger} \hat a_1- \hat a_1^{\dagger} \hat a_2\right),\\
\hat L_z&=\frac 12 \left(\hat a_1^{\dagger} \hat a_1-\hat a_2^{\dagger} \hat a_2\right),
\end{split}
\end{equation}
and $\hat L_\pm=\hat L_x\pm i \hat L_y$, it then follows that the action of the $\hat L_\alpha$ operators on the coherent state projector $\ket \Gamma \bra \Gamma$ can be represented by differential operators $\mathcal D(\hat L_\alpha)$, such that \cite{Narducci, Narducci2, Zhang}
\begin{equation}
\begin{split}
\hat L_+ \ket \Gamma \bra \Gamma&=\mathcal D(\hat L_+) \ket \Gamma \bra \Gamma\\
&=e^{i \phi} \left(\frac{N}{2}\sin \theta +\frac{i}{2} \tan \frac{\theta}{2} \partial_\phi - \sin^2 \frac{\theta}{2} \partial_\theta \right) \ket \Gamma \bra \Gamma,\\
\hat L_- \ket \Gamma \bra \Gamma&=\mathcal D(\hat L_-) \ket \Gamma \bra \Gamma\\
&=e^{-i \phi} \left(\frac{N}{2}\sin \theta -\frac{i}{2} \cot \frac{\theta}{2} \partial_\phi + \cos^2 \frac{\theta}{2} \partial_\theta \right) \ket \Gamma \bra \Gamma,\\
\hat L_z \ket \Gamma \bra \Gamma&=\mathcal D(\hat L_z) \ket \Gamma \bra \Gamma\\
&= \left(\frac{N}{2}\cos \theta +\frac{i}{2}  \partial_\phi - \frac 12 \sin \theta \partial_\theta \right) \ket \Gamma \bra \Gamma.
\label{eq:D_operators}
\end{split}
\end{equation}

In terms of these $\hat L_\alpha$ operators the Hamiltonian (\ref{eq:H}) can now be rewritten as
\begin{equation}
\hat H=-\frac{\Omega}{2}\left(\hat L_++\hat L_-\right)+U\left(\hat L_z^2+\frac{\hat N^2}{4}\right)+\Delta \hat L_z.
\label{eq:H_L}
\end{equation}
Using Eq.~(\ref{eq:Qdot}), Eq.~(\ref{eq:D_operators}) and Eq.~(\ref{eq:H_L}) we finally find
\begin{equation}
\begin{split}
\dot Q(\theta,\phi)&=i \tr \left(\mathcal D(\hat H) \ket \Gamma \bra \Gamma \hat \rho-\mathcal D(\hat H)^*\ket \Gamma \bra \Gamma \hat \rho\right)\\
&=-2 \imag \left[\mathcal D(\hat H)\right] \tr \left(\ket \Gamma \bra \Gamma \hat \rho \right)\\
&=\bigg[\frac{\Omega}{2}\cos(\phi)\left(\tan \frac{\theta}{2}-\cot \frac{\theta}{2}\right)\partial_\phi-\Omega \sin(\phi)\partial_\theta\\
& \qquad-UN\left(\cos \theta -\frac 1N\sin \theta \partial_\theta\right)\partial_\phi-\Delta \partial_\phi \bigg] Q(\theta,\phi)
\end{split}
\label{eq:Q_evolution}
\end{equation}
where $UN$ is of order one and the term containing second-order derivatives is therefore suppressed by a factor of $1/N$.

Using $\phi=q$, $\theta=\arccos(2p/N)$ we can also express Eq.~(\ref{eq:Q_evolution}) in $(q,p)$ as
\begin{equation}
\begin{split}
\dot Q(q,p)=&\Omega \sqrt{p_0^2-p^2} \sin(q) \frac{\partial Q(q,p)}{\partial p}\\
&-\left( 2Up+\Delta+\frac{\Omega p}{\sqrt{p_0^2-p^2}}\cos(q)\right) \frac{\partial Q(q,p)}{\partial q}\\
&-UN\left(\frac{p_0}{N}-\frac{p^2}{p_0 N}\right)\frac{\partial^2 Q(q,p)}{\partial q \partial p}.
\end{split}
\label{eq:Husimi}
\end{equation}
Comparing Eq.~(\ref{eq:Husimi}) to Eq.~(\ref{eq:Liouville}), we see that the evolution of the Husimi function is given by the classical Liouville equation plus a correction term containing second-order derivatives. Neglecting this last term leads to a ``Truncated Husimi approximation'' \cite{Witthaut2, Vermersch, Trimborn}, which, in analogy to the more familiar Truncated Wigner approximation \cite{Blakie_TWA, Polkovnikov_TWA, Sinatra_TWA}, can be understood as including quantum effects to first order. More specifically, quantum noise is modeled by sampling the initial conditions for the classical evolution from the quantum Husimi function, but quantum interference between different trajectories is neglected. Since $p$ and $p_0$ are of order $N$ and $UN$ is of order one, the quantum correction term seems to vanish in the classical limit ($N \to \infty$ with $UN$ held fixed), since the derivative with respect to $p$ comes with an additional factor of $1/N$.  One might therefore naively expect to recover the classical phase space evolution for large $N$. While this argument is often invoked, we will show in the following that this is not necessarily the case if a separatrix is involved in the classical evolution.

\subsection{Classical and quantum ergodization}
In the (semi-)classical description the finite return probability can be predicted accurately by Kruskal's theorem, as outlined in Sec.~\ref{semiclassical}. Since Kruskal's theorem makes statements about a phase space volume (more specifically: an energy shell) and how it is split up among multiple growing phase space regions, it is a crucial assumption in this application of Kruskal's theorem that the actual phase space density of the ensemble is uniform in this phase space volume, i.e. $\rho(q,p)=\rho(E)$ is an ergodic phase space distribution. In our case this means that one assumes that the energy shell in the lower separatrix lobe is filled with a uniform reduced density after the separatrix has been crossed. Of course this can only be true in some kind of approximate sense, because the phase space volume occupied by the ensemble is required to stay constant by Liouville's theorem. What actually happens is that as the initially uniform microcanonical ensemble crosses the separatrix and spills into its lower lobe, the exact phase space density rapidly develops an extremely fine ``swirling'' structure, in which the phase space volume which is actually occupied by the ensemble does indeed remain constant, but it is distributed within the larger phase space volume in extremely fine threads \cite{dimer}. The ``coarse-grained'' phase space density is thus uniform but reduced. For slower sweep rates the swirling becomes steadily finer, so that this approximation becomes perfect in the quasi-static limit.

It is not obvious how this ergodization mechanism could be realized in the quantum evolution, however. While the Husimi function does \emph{not} have to obey Liouville's theorem, so that ergodization might conceivably be even more effective quantum mechanically than it is classically, it turns out that quantum-classical correspondence breaks down precisely because of the swirling discussed above.

\subsubsection{Single initial quantum eigenstate}
\begin{figure}
\includegraphics[width=.45\textwidth]{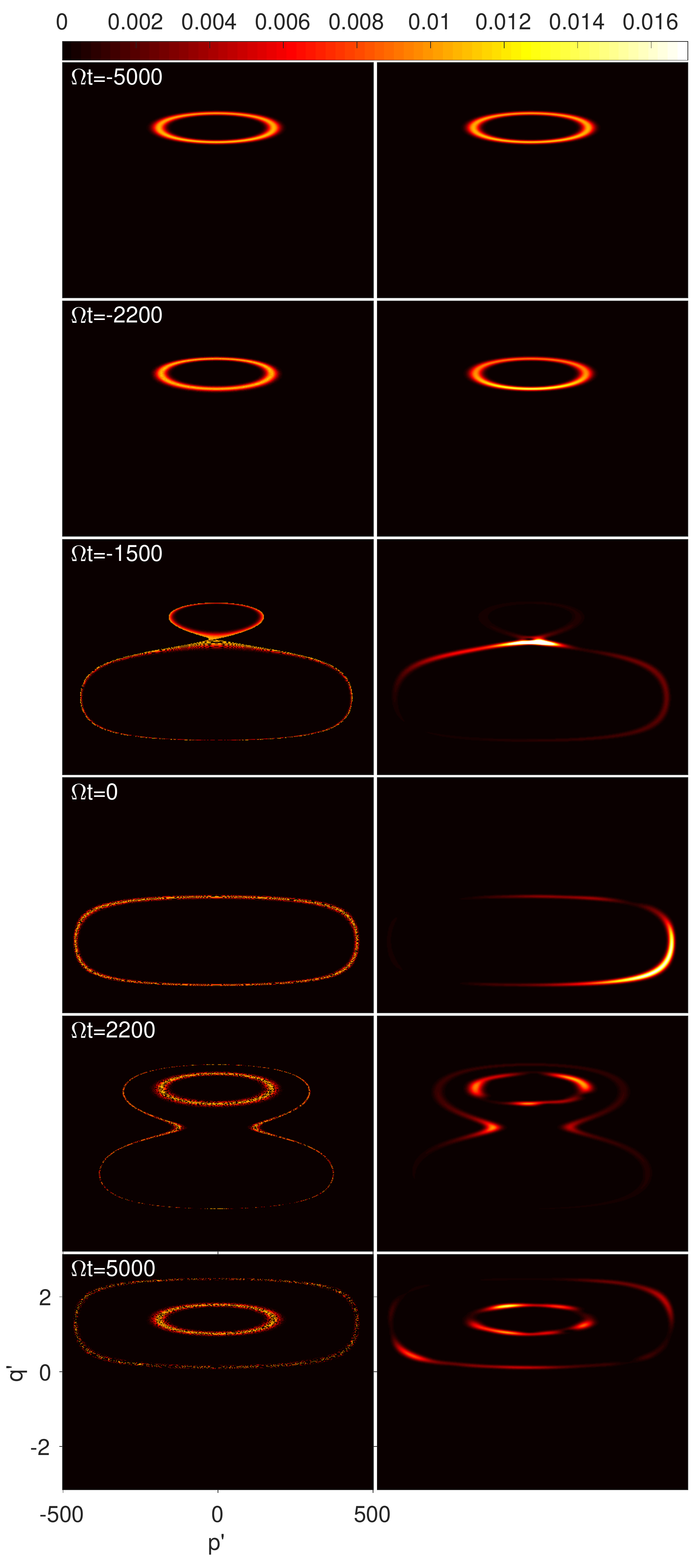}
\caption{Evolution of the Husimi function corresponding to a single initial state (37th adiabatic energy eigenstate, right) compared to the classical Liouville dynamics with the same initial phase space distribution (left). After the separatrix is crossed the two evolutions are quite different, despite the large total particle number. While the classical phase space density is ``grainy'' due to swirling (see text), the Husimi function is smooth, but far from ergodic, in that it has bright and dark spots. The system parameters are $N=1000$, $u=-3$, $\Delta_0/\Omega=-\Delta_I/\Omega=2$ and $T=5000\Omega^{-1}$.}
\label{fig:N=1000_single}
\end{figure}

Fig.~\ref{fig:N=1000_single} (right column) shows the evolution of the Husimi function for $N=1000$, $\Delta_0/\Omega=-\Delta_I/\Omega=2$, $T=5000\Omega^{-1}$ and a single energy eigenstate as the initial state (we have chosen the 37th eigenstate as in \cite{quantum_dimer_I}), along with the classical evolution of the same initial phase space density under Eq.~(\ref{eq:Liouville}) (left column). For the evolution of the Husimi function we solve the Schr\"odinger equation with the Hamiltonian (\ref{eq:H}) numerically and then calculate the Husimi function via Eq.~(\ref{eq:Husimi_def}). For the classical evolution we use the method of characteristics \cite{Courant, Sarra} to solve the Liouville equation, i.e. we evolve individual phase space patches (to which some probability is assigned by the initial conditions) under the Hamiltonian equations of motion. Note that every Husimi function is in principle a valid classical phase space density, since it is non-negative and normalizable, so that in particular the initial Husimi function is also a valid initial condition for Eq.~(\ref{eq:Liouville}). The corresponding classical evolution is then the Truncated Husimi approximation of the full quantum evolution. (Another reasonable choice for the classical initial phase space density would be a classical microcanonical distribution with energy boundaries between the 36th and 37th and 37th and 38th quantum energy eigenvalues. Since the initial Husimi function is already quite microcanonical this choice would lead to a very similar evolution.) For a better visual presentation we have again used the canonical coordinates that we used in Fig.~\ref{fig:classical_phase_space},
\begin{equation}
\begin{split}
q'&=\arctan\left(\frac{p}{\sqrt{p_0^2-p^2}\cos(q)}\right)\\
p'&=-\sqrt{p_0^2-p^2}\sin(q),
\end{split}
\end{equation}
which simply correspond to a rotation of the Bloch sphere before the mapping from $(\theta,\phi)$ to $(q,p)$ is performed. Note also that we have normalized the Husimi function according to $\int \mathrm d q' \mathrm dp' \; Q(q',p')=1$.

\begin{figure}
\centering
\subfloat[classical phase space density]{\includegraphics[width=.24\textwidth]{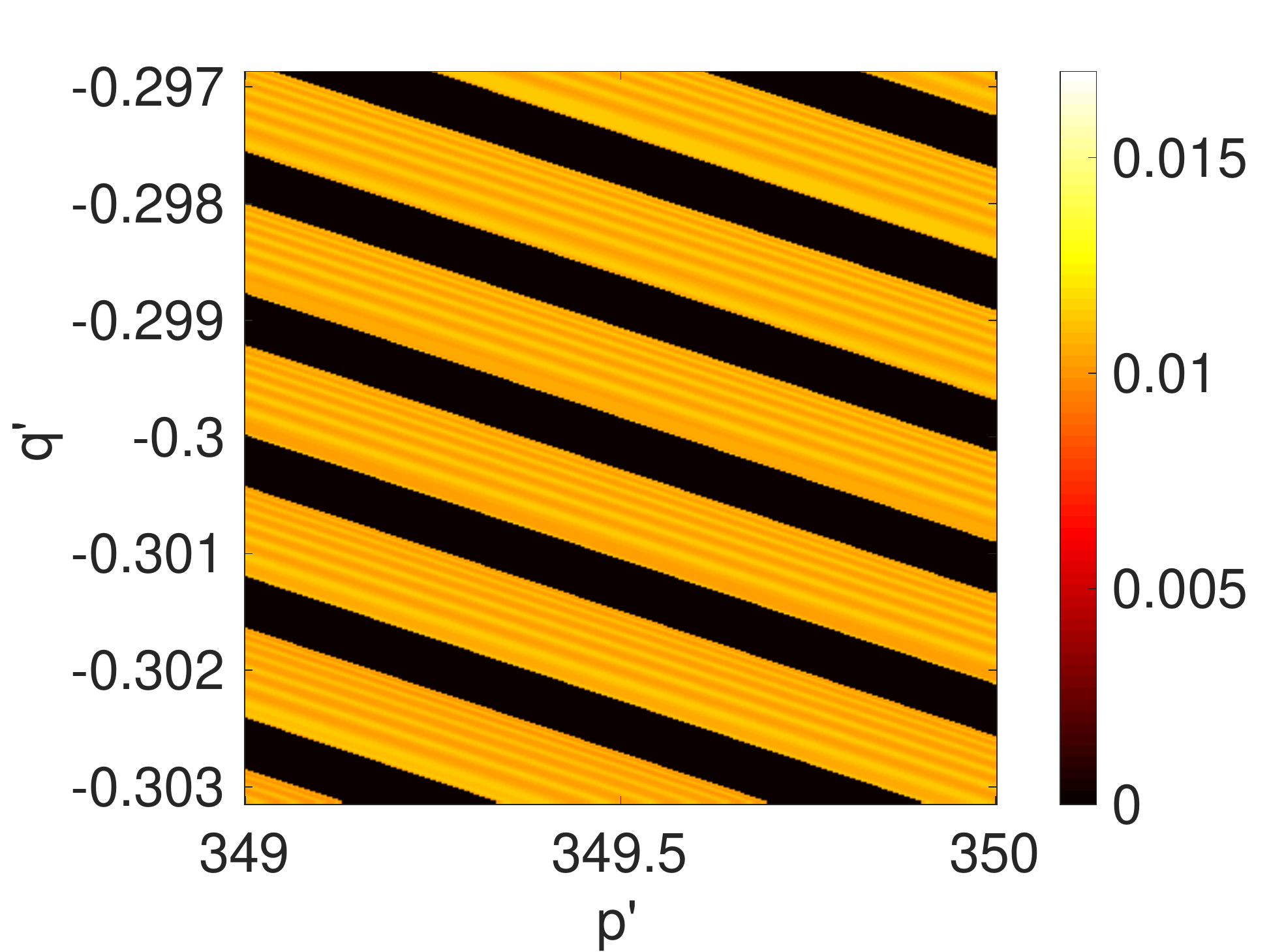}}
\subfloat[Husimi function]{\includegraphics[width=.24\textwidth]{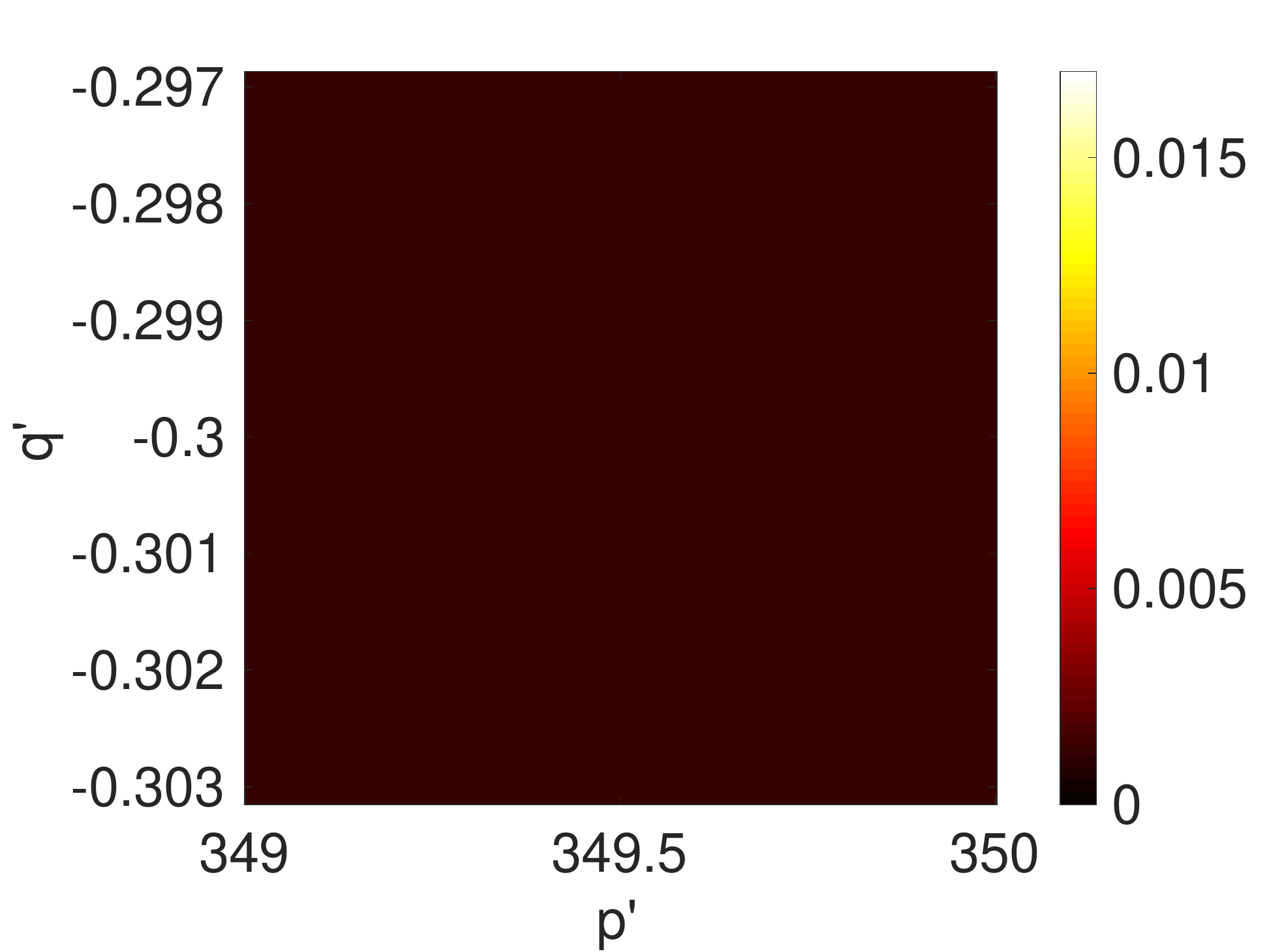}}
\caption{Higher resolution image of Fig.~\ref{fig:N=1000_single} at $t=0$ showing the swirling of the classical phase space density (a) and the Husimi function (b). The region shown here corresponds to approximately one pixel in Fig.~\ref{fig:N=1000_single} and is much smaller than $\hbar$.}
\label{fig:swirling}
\end{figure}

Before the separatrix is crossed around $\Omega t=-1500$, the two evolutions are very similar, as expected, because the quantum correction term is suppressed by $1/N$. After the separatrix has been crossed the classical phase space density spreads almost uniformly into a larger phase space volume along an energy contour that is determined by the initial action \cite{dimer}. The classical phase space density actually has the extremely fine swirling structure mentioned above, but this structure is not fully resolved in our simulation with a finite sampling of phase space and therefore appears as seemingly random black dots spread through the energy shell. The swirling is so fine that a phase space volume much smaller than the Heisenberg limit $\hbar$ ($=1$ in our units) has to be resolved to reveal it. Fig.~\ref{fig:swirling} shows a part of the panels with high resolution, at the end of the forward sweep ($t=0$).
Note that the phase space area shown in this figure is $0.006 \hbar$ and corresponds to approximately one pixel in Fig.~\ref{fig:N=1000_single}. For slower sweep rates the swirling becomes even finer, approaching a uniform but reduced phase space density, as discussed above. Consequently the swirling mechanism is responsible for (coarse-grained) ergodization. We emphasize again that this ergodization mechanism is crucial for the explanation of irreversibility in the classical limit, since Kruskal's theorem assumes an ergodic phase space distribution. Details can be found in \cite{dimer}. The fineness of the swirling also demonstrates why, even though the classical evolution is deterministic, we speak of \emph{probabilistic} hysteresis: to guarantee a reversible evolution the initial conditions would have to be tuned to be within phase space volumes of comparable size to the very fine scale of the swirling. For the sweep rate and particle number presented here this would mean controlling the initial phase space location of the system on scales much smaller than the Heisenberg limit. Since this is clearly impossible, experiments would show effectively random run-to-run alternations between the two final outcomes even if the evolution were classical.

The Husimi function, on the other hand, while also being localized on an energy contour, does not spread uniformly along this contour. Instead it forms a rapidly changing, more or less localized pattern (see also Fig.~\ref{fig:entropy}), so that there is \emph{no} ergodization. In the Landau-Zener picture \cite{quantum_dimer_I} (where the sweep is considered as a series of Landau-Zener crossings) this is an interference effect of the many involved adiabatic eigenstates. In the phase space picture, however, the classical ergodization mechanism via fine swirling breaks down despite the large particle number. The reason for this is that as soon as the classical swirling structure even begins to develop, the second derivatives in the quantum correction term in Eq.~(\ref{eq:Husimi}) become extremely large, because the swirling introduces steep gradients in $Q$ between the high and low probability stripes, as shown in Fig.~\ref{fig:swirling}. Therefore the same swirling mechanism that leads to ergodization of the classical phase space density is also directly responsible for the breakdown of quantum-classical correspondence. What turns out to happen quantum mechanically is that the sub-$\hbar$ sized swirling structure does not develop in the Husimi function. 

With the breakdown of the classical ergodization mechanism there is no reason to expect ergodization of the Husimi function, and it is indeed absent as we have confirmed numerically. The observed localization of the Husimi function is a purely quantum phenomenon and has dramatic effects on the evolution in phase space, despite the large particle number $N$ that would at first sight suggests good quantum-classical correspondence. 

\begin{figure}
\centering
\includegraphics[width=.45\textwidth]{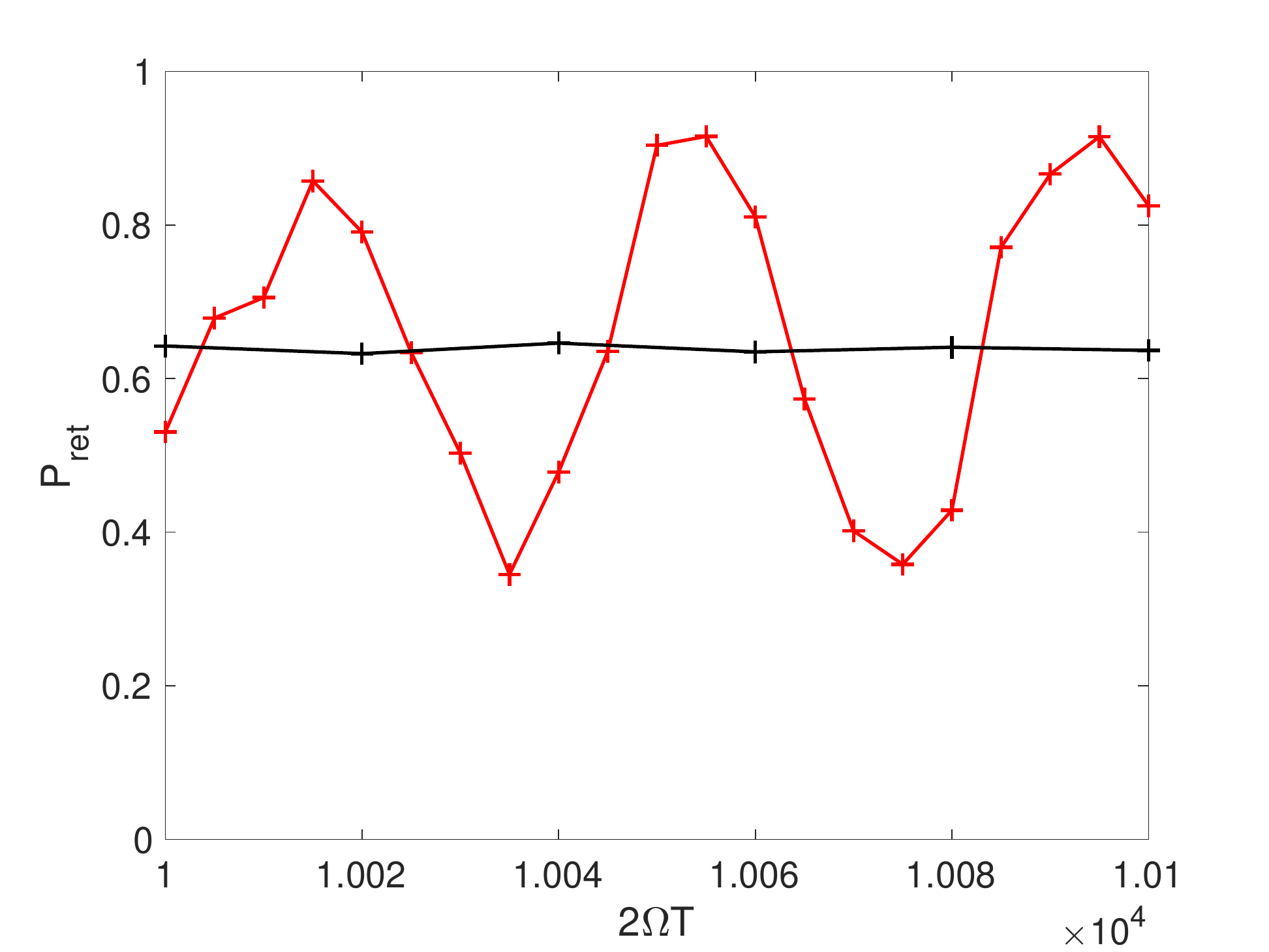}
\caption{Dependence of the return probability $P_{\mathrm{ret}}$ on the total sweep time $2 T$ for $N=1000$, $u=-3$, $\Delta_0/\Omega=-\Delta_I/\Omega=2$ and a single initial state. The red line shows the quantum return probability if the initial state is the 37th eigenstate of the initial Hamiltonian. The black line shows the corresponding classical return probability for an initial phase space density equal to the initial Husimi function. The slight variation of the classical return probability is due to our finite phase space resolution (sampling error).}
\label{fig:p_N=1000_single}
\end{figure}

During the backward sweep both the classical phase space density and the Husimi function split into two well separated parts, corresponding to the returning and non-returning fraction. However, due to the oscillatory behavior of the Husimi function the return probability $P_{\mathrm{ret}}$ at the end of the sweep depends sensitively on the sweep rate, in contrast to the classical return probability, see Fig.~\ref{fig:p_N=1000_single}. The return probability in the quantum case is defined in analogy to the classical return probability as the phase space integral of the Husimi function over the inner ring in Fig.~\ref{fig:N=1000_single} at $t=T$. This integral defines the return probability unambiguously, quantum mechanically as well as classically, because in both cases the inner and outer rings are well separated for large $N$. 

In \cite{quantum_dimer_I} it was shown that the forward sweep leads to a superposition of adiabatic eigenstates. In the phase space picture this superposition is responsible for the rapid dynamics of the Husimi function. Since the return probability depends on the shape and localization of the Husimi function when it crosses the separatrix, the superposition of adiabatic eigenstates generally leads to a non-trivial dependence of the return probability on the total sweep time. The almost periodic dependence shown in Fig.~\ref{fig:p_N=1000_single} is due to the fact that only a relatively small number of adiabatic eigenstates, within a narrow range of adiabatic energies, have an appreciable amplitude. With weak nonlinearity in our system, the adiabatic eigenfrequencies of these superposed states are nearly even spaced within their narrow energy range, and thus nearly commensurate, so that the periodic maxima and minima of the return probability as a function of sweep rate are in fact similar to simple two-state Ramsey fringes.

\subsubsection{Initial ensemble of quantum eigenstates}
\begin{figure}
\includegraphics[width=.45\textwidth]{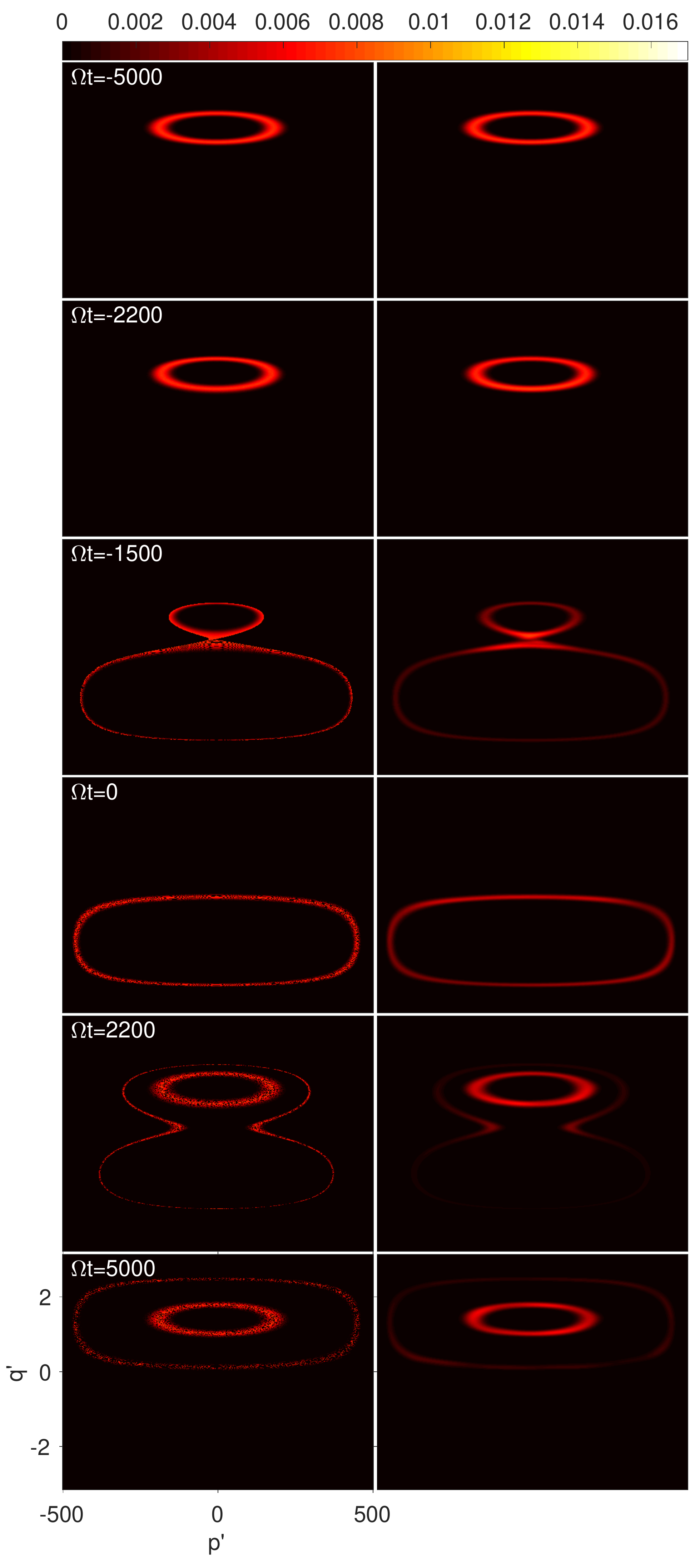}
\caption{Evolution of the Husimi function corresponding to a microcanonical ensemble of 20 initial states (right) compared to the classical Liouville dynamics with the same initial phase space distribution (left). The oscillations of the Husimi function seen in Fig.~\ref{fig:N=1000_single} are strongly suppressed, so that we observe much better quantum-classical correspondence in the return probability. Parameters are the same as in Fig.~\ref{fig:N=1000_single}. The 20 quantum states are the initial 28th to 47th states, so that the mean energy of the ensemble is essentially the same as in Fig.~\ref{fig:N=1000_single}.}
\label{fig:N=1000_ensemble}
\end{figure}

If we start with a microcanonical ensemble of quantum states initially, instead of with a single energy eigenstate, we obtain the evolution displayed in Fig.~\ref{fig:N=1000_ensemble}. Our ensemble contains 20 consecutive initial adiabatic eigenstates, chosen in such a way that the mean energy of the ensemble is essentially the same as the energy of the single state in Fig.~\ref{fig:N=1000_single}.
Because the energy width is still small due to the large particle number, which makes the spacing between quantum energy eigenstates small enough that 20 states is a narrow range of eigenvalues, the classical evolution of this ensemble is almost indistinguishable from the classical evolution in Fig.~\ref{fig:N=1000_single}. The evolution of the Husimi function before the separatrix is crossed also remains very similar to the evolution shown in Fig.~\ref{fig:N=1000_single}. 

For even this narrow 20-state microcanonical ensemble, however, the oscillations of the Husimi function that were found for a single initial state after the separatrix had been crossed are now suppressed, and ergodization is restored to a good approximation. The classical swirling structure is also present in the classical case with the larger initial energy width, and this still destroys naive quantum-classical correspondence as explained above, by inducing a large quantum correction term. In the 20-state quantum ensemble, however, a new quantum ergodization mechanism has emerged. The Husimi function of a mixed state is simply the weighted sum of the Husimi functions of the pure states that have been mixed (recall the definition of the Husimi function Eq.~(\ref{eq:Husimi_def})). For our microcanonical ensemble the Husimi function is therefore the average of many Husimi functions like the one shown in Fig.~\ref{fig:N=1000_single}. In each of these Husimi functions the localized dark and bright patches at any given time appear at different locations that depend strongly on energy. Averaging over energy therefore averages out the bright and dark patches, yielding an evenly ergodic total Husimi function. 

\begin{figure}
\centering
\includegraphics[width=.45\textwidth]{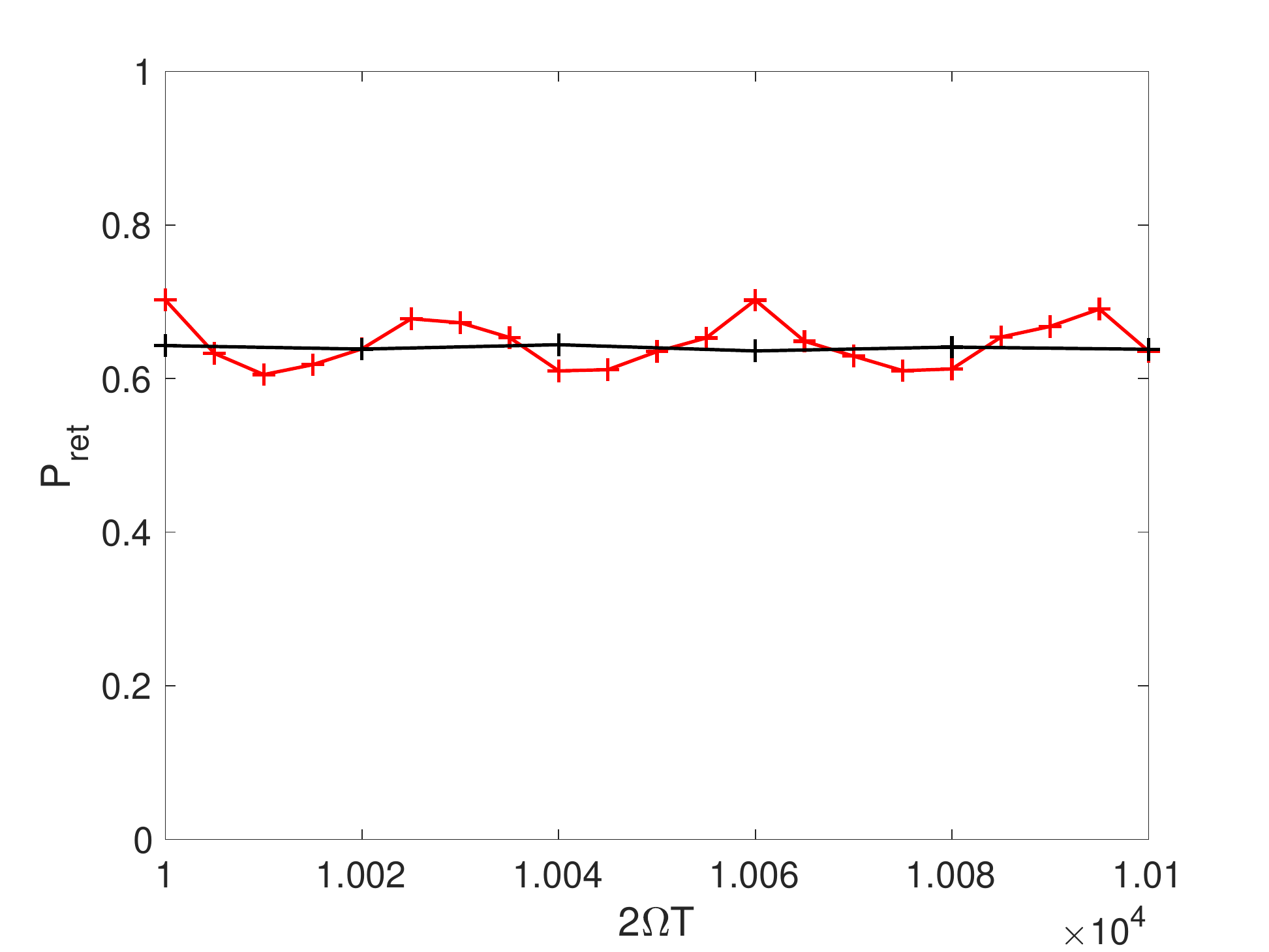}
\caption{Dependence of the return probability $P_{\mathrm{ret}}$ on the total sweep time $2 T$ for $N=1000$, $u=-3$, $\Delta_0/\Omega=-\Delta_I/\Omega=2$ and an ensemble of 20 initial states (28th to 47th energy eigenstates) with mean energy similar to the 37th state (red line). The black line shows the corresponding classical return probability for an initial phase space density equal to the initial Husimi function. The oscillations of the return probability are much smaller than in Fig.~\ref{fig:p_N=1000_single} and are expected to vanish completely for larger initial energy width.}
\label{fig:p_N=1000_ensemble}
\end{figure}

Once quantum ergodization is established, the quantum evolution of the finite-width ensemble shows much better agreement with the semiclassical phase space evolution, because now the quantum correction term in Eq.~(\ref{eq:Husimi}) really does remain small (of order $1/N$). In particular the return probability loses its high sensitivity to the sweep rate and approaches the classical value, see Fig.~\ref{fig:p_N=1000_ensemble}. With no fine swirling, the quantum evolution of the Husimi function is approximately Liouvillian for large $N$, and so Kruskal's theorem then also holds approximately for the Husimi function when the separatrix is encountered during the backward sweep and the return probability is determined.

It is therefore clear that the classical limit is not obtained simply by letting $N\to \infty$ with a single quantum state, but an ensemble with finite energy width is needed for good quantum-classical correspondence of the return probability in probabilistic hysteresis, as was also found in \cite{quantum_dimer_I}. In the quantum phase space formalism this can be explained by the fact that an ensemble containing enough quantum states effectively smears out the localized individual Husimi functions, so that the total Husimi function becomes effectively ergodized like the classical phase space density, albeit for quite different reasons. For their different reasons, the finely swirled exact classical phase space density and the energy-averaged Husimi function both behave very similarly to a smooth ergodized phase space density. The Husimi function for sufficient initial energy width and the classical phase space density thus effectively behave very similarly to each other, up to the small quantum-classical discrepancies of order $1/N$ that one naively expects from the Liouville and Husimi evolution equations.

\subsection{Entropy}
In a macroscopic system ergodization leading to irreversibility is associated with the growth of entropy. Microscopically, however, the phase space volume that is occupied by a classical ensemble is invariant under Hamiltonian time evolution---and so is the entropy. This is a direct consequence of Liouville's theorem, which is often expressed in the statement that classical phase space flow is like the flow of an incompressible fluid. The entropy in which one is normally interested, however, is some coarse-grained entropy, which is usually defined in reference to a limited resolution in phase space or to some implicit time-averaging. We realize this coarse-graining at every instant by time-averaging the fine-grained phase space density to obtain the coarse-grained density $\rho_c$
\begin{equation}
\rho_c(q',p')=\lim_{\widetilde T \to \infty} \frac{1}{\widetilde T} \int_0^{\widetilde T} \mathrm dt \; \rho(q',p';t),
\label{eq:rho_cg}
\end{equation}
where the time dependence of the phase space density on the right side is due to the evolution under the frozen Hamiltonian with fixed $\Delta$. Note that if the system is described in action-angle coordinates this provides coarse-graining in the angle coordinate only, but not in the action (or, equivalently, energy). Coarse-graining thus smears out the fine swirling structure found in the classical evolution, so that the coarse-grained entropy should increase when the separatrix is crossed and swirling occurs.

\begin{figure}
\includegraphics[width=.45\textwidth]{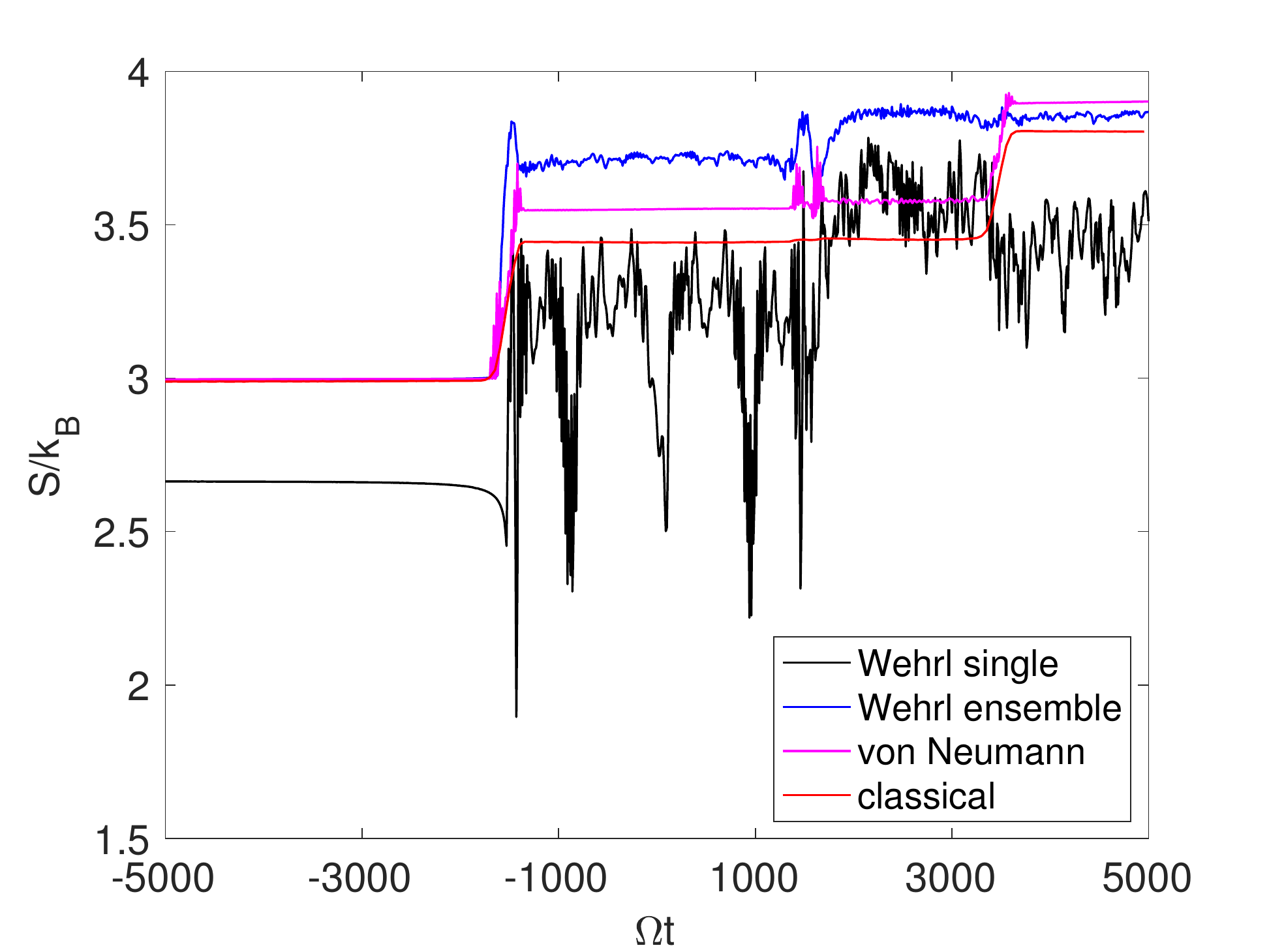}
\caption{Wehrl entropy $S_W$ corresponding to the quantum evolutions shown in Fig.~\ref{fig:N=1000_single} (black) and Fig.~\ref{fig:N=1000_ensemble} (blue).  While the oscillation of the Husimi function in the case of a single initial state leads to a strong oscillation of the Wehrl entropy, the oscillations of the Wehrl entropy in the case of an initial ensemble of states are suppressed. The red line shows the coarse-grained entropy of the classical simulation of Fig.~\ref{fig:N=1000_ensemble} for comparison. For further comparison, the magenta line shows the von Neumann entropy of the time-averaged quantum density matrix (see text). Since the definitions of the Wehrl and classical entropies allow an arbitrary constant shift from the phase space measure, we have used this to set all three entropies equal for the mixed initial state.}
\label{fig:entropy}
\end{figure}

In the quantum system, on the other hand, the von Neumann entropy remains constant, since we do not trace out any degrees of freedom and the evolution is unitary. Therefore the question arises: what quantum entropy corresponds to the classical coarse-grained entropy? As we have already seen there is a close analogy between the classical coarse-grained phase space density and the Husimi function, so that one natural quantum analogue of the coarse-grained entropy is the \emph{Wehrl entropy} \cite{Wehrl,Wehrl2}
\begin{equation}
S_W=-k_B \int \mathrm dq' \mathrm dp' \; Q(q',p') \log\left[Q(q',p')\right].
\end{equation}
Because the flow of the Husimi density in phase space is \emph{not} incompressible, this entropy \emph{can} increase during the evolution, even without coarse-graining.  Fig.~\ref{fig:entropy} shows the Wehrl entropy for the two cases discussed above, of a single initial eigenstate (black) and a 20-state microcanonical ensemble (blue). Note that the Wehrl entropy for the ensemble of states is always higher than the Wehrl entropy for the single quantum state, simply because the initial Husimi function is wider.
In the case of a single initial state (black), where there is no quantum ergodization, the Wehrl entropy can be used to quantify the localization of the Husimi function. The strong and fast oscillations shown in Fig.~\ref{fig:entropy} for $t\gtrsim -1500 \Omega^{-1}$ therefore confirm what we have already seen for a few discrete values of $t$ in Fig.~\ref{fig:N=1000_single}: fast ``collapse and revival'' of the Husimi function instead of smooth ergodization.

In the case of the ensemble of initial states (blue) these oscillations of the Wehrl entropy are much smaller, reflecting the smoothness of the Husimi function observed in Fig.~\ref{fig:N=1000_ensemble}. Does this Wehrl entropy then correspond to the classical coarse-grained entropy? To answer this question we also show the coarse-grained classical entropy, obtained from the classical coarse-grained phase space density Eq.~(\ref{eq:rho_cg}) of Fig.~\ref{fig:N=1000_single}, as a red line in Fig.~\ref{fig:entropy}. The red and blue lines agree initially, until the separatrix crossing, because as long as there is no swirling quantum and classical evolution correspond closely at this $N$, and because without swirling of the initially ergodic classical ensemble, coarse-graining has no effect on it.

The classical entropy (red) increases in two steps and stays essentially constant in between. The first step occurs around $\Omega t=-1500$ when the separatrix is crossed during the forward sweep and the coarse-grained density fills the larger phase space volume in the growing lower separatrix lobe. The second step around $\Omega t=3500$ is due to the same mechanism: After $\Omega t=2500$, where $\Delta$ is negative again, the phase space region outside the separatrix is shrinking. This means that the outer shell, representing the non-returning fraction, crosses the separatrix again when it makes the transition from the figure-eight shape shown in the second-to-last panel of Fig.~\ref{fig:N=1000_ensemble} to the ellipsoidal shape shown in the last panel. In the same way as when the separatrix was crossed during the forward sweep, the outer shell merges with another empty energy shell (which is leaving the lower separatrix lobe), so that the phase space density decreases and entropy increases. This transition does not influence the return probability, because which trajectories return to the initial state has already been decided much earlier and this phase space dilution process involves only the part of the phase space density that did not return, anyway.

The Wehrl entropy for the ensemble of states does not fully agree with the classical entropy, but shows additional features that are due to its finite minimum width. When the separatrix is crossed around $\Omega t=-1500$ the Wehrl entropy increases to a value above the classical entropy, simply because the Husimi function for our large but finite $N$ is a little wider than the classical phase space density (see Fig.~\ref{fig:N=1000_ensemble}). When the separatrix is encountered again during the backward sweep around $\Omega t=1500$, the classical phase space density and the Husimi function split into two parts, the outer figure-eight shaped shell and the inner ellipsoidal shell. While the classical occupied phase space volume stays constant during this transition, so that the entropy also does not change, the outer shell becomes much thinner than the Husimi function can ever be. The greater width of the Husimi function compared to the classical phase space distribution then leads to a small additional increase of the Wehrl entropy that has no classical counterpart. In general a finite width Husimi function will always have a larger (Wehrl) entropy than the corresponding classical distribution, because the Husimi function has blurred edges in comparison to the classical phase space density. This also means that the Wehrl entropy can decrease if the shape of the classical distribution is deformed in a such a way that the length of the edges decreases. This is the case when the outer shell crosses the separatrix around $\Omega t=3500$, and apparently the decrease of the Wehrl entropy in this process is almost compensated by the increase of the entropy that was expected from the classical considerations.

While the Wehrl entropy for our large but still finite particle number $N$ does therefore not fully agree with the classical entropy, a somewhat better agreement can be found for the alternative entropy
\begin{equation}
S=-k_B\sum_n p_n \log(p_n)
\label{eq:cg_vN_entropy}
\end{equation}
with
\begin{equation}
p_n=\braket{n|\hat \rho|n}
\end{equation}
where $\ket n$ are the adiabatic eigenstates. This entropy, which was introduced by von Neumann \cite{vonNeumann} for the case of a pure quantum state, can be understood as the more widely known mixed-state von Neumann entropy $-k_B \mathrm{Tr}(\hat{\rho}\ln\hat{\rho})$ of the time-averaged density matrix \begin{equation}
\hat \rho_c=\lim_{\widetilde T\to \infty} \frac{1}{\widetilde T}\int_0^{\widetilde T} \mathrm dt \; \hat \rho(t),
\end{equation}
where $\hat \rho(t)$ on the right side is again given by the evolution under the frozen Hamiltonian with fixed detuning $\Delta$.  The off-diagonal terms in the density matrix oscillate because the quantum phases of the adiabatic eigenstates evolve at different speeds, and because the phase difference between different adiabatic eigenstates $\ket n$, $\ket m$ is thus effectively random, the off-diagonal terms are averaged out. In fact, due to the slowness of the sweep compared to the time scale on which the adiabatic phases change, the averaging can also be done for finite $\widetilde T$ and $\hat \rho(t)$ given by evolution under the actual time-dependent Hamiltonian. The same reasoning was the motivation for the incoherent Landau-Zener approximation in \cite{quantum_dimer_I}.

The entropy Eq.~(\ref{eq:cg_vN_entropy}) is thus conceptually similar to the classical coarse-grained entropy, where time-averaging smeared out the classical swirling structure. It and its generalizations to other bases besides energy eigenstates have been shown to satisfy analogs of the Boltzmann $H$-theorem \cite{Han,Hu} and have been proposed elsewhere as useful tools for analyzing quantum-classical correspondence \cite{Fang}. Since the entropy Eq.~(\ref{eq:cg_vN_entropy}) may therefore also be considered as an alternative quantum entropy, we plot it in Fig.~\ref{fig:entropy}; it shows better agreement with the classical coarse-grained entropy compared to the Wehrl entropy. Note that the classical and Wehrl entropies in Fig.~\ref{fig:entropy} have both been shifted by a fixed amount, as one is always free to do by changing the size of the elementary phase space cell, so that the blue, red and magenta curves start out at the same value of $S=k_B\log(20)$, since we start with a microcanonical ensemble of 20 quantum states. The von Neumann entropy of the time-averaged density matrix then increases whenever the evolution of one of the 20 initial states leads to a superposition of multiple adiabatic eigenstates, which can, for large $N$, only happen close to where the separatrix is crossed classically, as has been discussed in \cite{quantum_dimer_I}. We may therefore say that the quantum analogue of classical swirling is the quantum superposition of adiabatic eigenstates.

To summarize, in the case of sufficient initial energy width corresponding patterns of plateaus and jumps can be seen in the quantum and classical entropies. Precise agreement between quantum and classical entropies even in the large-$N$ limit is hard to confirm, however. This is partly due to the basic fact, ultimately due to the uncertainty principle, that probability density in phase space is just not really well-defined in quantum mechanics.  Whatever kind of quantum quasi-probability function one may define in phase space, the width of this cannot be guaranteed to coincide exactly with the width of any classical phase space density, which can become arbitrarily thin. In particular the Husimi function of a given quantum state cannot be considered to represent the probability distribution in phase for that quantum state, since integrating the Husimi function over position or momentum will in general not yield the correct probability distribution of the remaining canonical observable in that state. It is still true that every quantum Husimi function is valid as \emph{a} classical ensemble in phase space, but the Husimi function of a quantum microcanonical ensemble is \emph{not} a classical microcanonical ensemble, and this makes the comparison between the Wehrl entropy and the entropy of the time-averaged density matrix difficult. Even though correspondence of the return probabilities is restored, quantum-classical correspondence is thus not perfect, even at large $N$ and sufficient energy width for quantum ergodization. 

Besides the dramatic failure of quantum-classical correspondence for large particle numbers and a single initial state, there is also the more expected failure at low particle numbers. In particular the adiabatic limit in the quantum system is very different from the classical adiabatic limit, in that the evolution is reversible in the former limit whereas irreversibility persists in the latter limit. We will show in the next section how the quantum adiabatic limit appears in the Husimi phase space formulation.

\section{Reversibility by Macroscopic Quantum Tunneling}
After having demonstrated how the classical limit of the return probability can emerge from the quantum phase space description for large particle numbers, sufficient energy width, and very but finitely slow sweep rate, we now consider the case of ultimately slow sweep rates, in which the quantum adiabatic limit of completely reversible evolution is always different from the classical quasi-static limit of probabilistic hysteresis. As has been shown in \cite{quantum_dimer_I}, even for quite small $N$ this quantum adiabatic limit requires quite unrealistically slow sweep rates, because the energy gaps with respect to which the sweep has to be adiabatic are exponentially small in $N$. For $N=10$, however, we can at least reach the quantum adiabatic limit with a numerical simulation, in the sense that the system remains in the same adiabatic eigenstate with probability $>0.99$. We can achieve this for the same Hamiltonian parameters as in Fig.~\ref{fig:N=1000_single} with $T=10^8 \Omega^{-1}$. For $N=20$ and the same Hamiltonian parameters, on the other hand, the total sweep time $2T$ already has to be on the order of $10^{15} \Omega^{-1}$ to obtain a fully reversible evolution, which would be around 30 years if $\Omega$ were in the experimentally typical MHz regime, or even 30000 years for $\Omega$ in the experimentally feasible kHz regime. The question of why the quantum adiabatic limit is so insanely hard to reach for higher $N$ can actually be answered by the numerically achievable case of $N=10$.

\begin{figure}
\centering
\includegraphics[width=.45\textwidth]{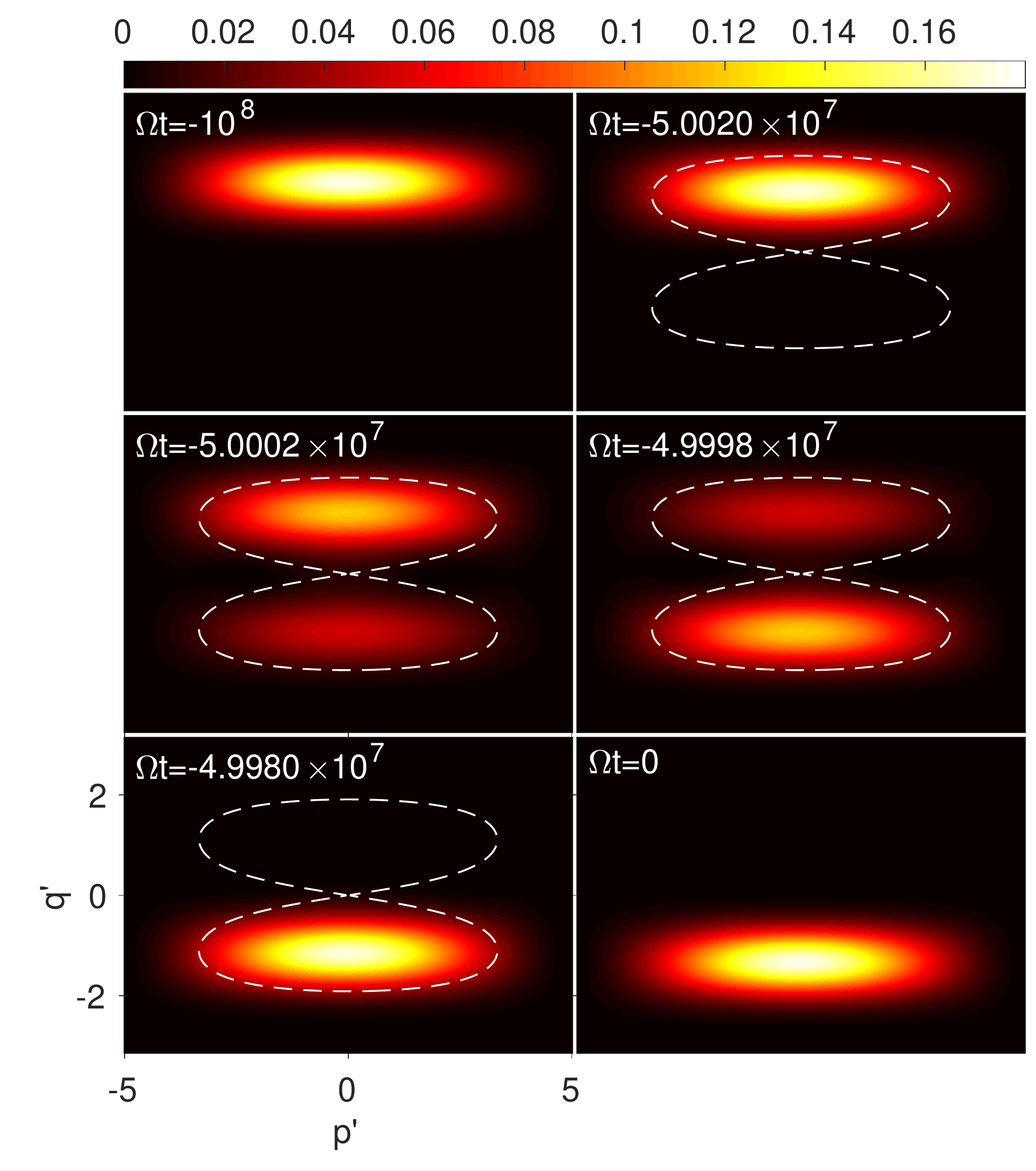}
\caption{Evolution of the Husimi function in the reversible quantum adiabatic limit for $N=10$, $T=10^8 \Omega^{-1}$, $u=-3$ and $\Delta_0/\Omega=-\Delta_I/\Omega=2$ during the forward sweep, where the initial state is the ground state. The panels for the backward sweep are essentially identical to the panels shown, but in reverse order. Reversibility in this extreme case is restored by macroscopic quantum tunneling: All atoms tunnel collectively through the separatrix, which is indicated by the dashed white line, in the forward and backward sweep. Accordingly, the Husimi function also tunnels through the separatrix: instead of continuously flowing through the separatrix as in Fig.~\ref{fig:N=1000_single} and Fig.~\ref{fig:N=1000_ensemble}, the Husimi function simply fades away on one side of the separatrix and grows on the other side, without ever having visible support on the separatrix itself. Because the system always stays in a single adiabatic eigenstate, interference effects like those in Fig.~\ref{fig:N=1000_single} are absent.}
\label{fig:Husimi_N=10}
\end{figure}

Fig.~\ref{fig:Husimi_N=10} shows the evolution of the Husimi function for $N=10$ and $T=10^8 \Omega^{-1}$, where the initial state is the ground state at $\Delta_I/\Omega=-2$, meaning that initially almost all atoms are in the first mode and the Husimi function is localized in the upper half of the phase space $q'>0$. 
We find that around $\Delta\sim 0$ ($t\sim -T/2$), where the energy difference between the two lowest energy eigenstates becomes minimal, the Husimi function \emph{tunnels} into the lower half of the phase space. At this time most of the corresponding classical orbits would still be far away from the separatrix and would therefore stay in the upper separatrix lobe, until a much larger detuning is reached and they are touched by the shrinking separatrix lobe. This tunneling through an energetic barrier (and through the separatrix in phase space) is an example of \emph{macroscopic quantum tunneling} \cite{Owerre, Leggett, Takagi, Leggett2}: in the adiabatic limit all atoms tunnel from the first mode to the second mode collectively during the forward sweep around $\Delta\sim 0$. During the backward sweep the same tunneling occurs again, so that the evolution is reversible. Because the energetic barrier is very high, this macroscopic tunneling is so extremely slow that it only plays a role for extremely slow sweep rates. For the classical phase space distribution, in contrast, tunneling through an energetic barrier is not possible at all, no matter how slow the sweep may be. This tunneling through a separatrix, which is only possible quantum mechanically, is the reason for the incommutability of the semiclassical and adiabatic limits.

Because \emph{all} particles have to tunnel through the barrier it is also clear that the sweep time needed to reach the adiabatic limit quickly increases with $N$. For higher initial energy the energetic barrier is lower and a smaller number of particles has to tunnel ($N-2(i-1)$ for the initially $i$-th state). As long as the total particle number is not very small, however, the effect of macroscopic tunneling can still be neglected for realistic sweep rates until the Husimi function comes close to the separatrix and the energetic barrier is very low. In this case the only practical consequence of macroscopic tunneling is that the crossing of the separatrix happens in a slightly larger $\Delta$ range than would be expected classically.

\section{Conclusion}
In conclusion we have provided a phase space description of how irreversibility in the form of probabilistic hysteresis occurs in a dissipationless quantum system. In particular we have used the Husimi function to show that the quantum evolution is closely related to the classical evolution, with the usual $1/N$ dependence of the quantum correction term, but in spite of this scaling, which naively suggests good quantum-classical correspondence for large $N$, we have found that the particular ergodization mechanism due to ``swirling'' in the classical evolution leads to a breakdown of  quantum-classical correspondence. This inhibits quantum ergodization of the Husimi function, precisely at the point where classically the separatrix is crossed and irreversibility begins its onset. This quantum lack of ergodization has the result that the quantum return probability oscillates around the semiclassical value if the sweep time is varied, even for very large $N$. 

The classical limit for the return probability thus only emerges in the full quantum evolution if there is a specifically quantum ergodization mechanism. Such an ergodization mechanism appears naturally if an ensemble of initial quantum states is considered instead of a single state, because the Husimi function in this case is the superposition of the Husimi functions of the individual pure states, each of which is quite localized but rapidly oscillating. The classical return probability in probabilistic hysteresis is determined under Kruskal's theorem by the combination of Liouvillian evolution and effective ergodization, and so the classical return probability is only recovered from the quantum evolution if, besides the usual mean-field limit $N\to \infty$, a finite initial energy width is also allowed, since in this case Kruskal's theorem can also be applied to the Husimi evolution to a good approximation.

Even for large $N$ and sufficient initial energy width for quantum ergodization, quantum-classical correspondence is not perfect, though. While to the naked eye the Husimi function and the classical phase space density appear almost indistinguishable in this case, the comparison of the Wehrl entropy and the classical coarse-grained entropy reveals that the Husimi function still has distinctive quantum features. In particular the Husimi function always has a finite width due to the uncertainty principle, so that even at large $N$ it can be considerably wider than the corresponding classical phase space density. Good correspondence between the quantum and classical return probabilities still appears for finite $N$, however, because the classical return probability for thin energy shells only depends weakly on energy itself, so that the additional width of the Husimi function does not change the return probability significantly. It remains an open question whether the entropy discrepancies due to thinness eventually vanish in the true classical limit of $N \to \infty$ or for larger initial energy width. After all, the Husimi function cannot be interpreted as a true probability distribution in phase space.

In the opposite limit of small particle numbers we have shown that it is macroscopic quantum tunneling that is responsible for the different behavior of the quantum and classical systems in the quasi-static limit of infinitely slow sweep rate. Unlike the classical system, the quantum system can tunnel through the separatrix---and can thereby remain in the same adiabatic energy eigenstate throughout the whole forward-and-back sweep cycle, so that reversibility is restored. Since the energetic barrier is high, however, and the number of atoms that have to tunnel is of order $N$ for not too high initial energy, this tunneling is so very slow that for realistic sweep rates it plays no role unless the total particle number is very low ($N\lesssim 20$).

In final summary, we have offered a complementary viewpoint on quantum probabilistic hysteresis to the description in terms of Landau-Zener crossings that was obtained in \cite{quantum_dimer_I}. We have identified why quantum-classical correspondence breaks down in probabilistic hysteresis (fine classical swirling) and why the classical and quantum adiabatic limits are different (macroscopic quantum tunneling). Furthermore the phase space picture that we have presented in this paper leads us at least to the conjecture that the exact unitarity of quantum evolution, which is a close analog to the Liouvillian incompressibility of classical phase space flow, must be a fundamentally robust point of correspondence between quantum and classical dynamics, such that given some form of ergodization in either kind of dynamics, conclusions like those of the classical Kruskal's theorem must emerge in both cases. A fully quantum analog of Kruskal's theorem thus becomes a natural goal for future study; it might be approached by returning to the Landau-Zener picture of our previous paper \cite{quantum_dimer_I}.

\acknowledgements
The authors acknowledge support from State Research Center OPTIMAS and the Deutsche Forschungsgemeinschaft (DFG) through SFB/TR185 (OSCAR), Project No. 277625399.

%

\end{document}